\documentclass[a4paper]{jpconf}
\usepackage{graphicx}
\def\be{\begin{equation}}
\def\ee{\end{equation}}
\def\bea{\begin{eqnarray}}
\def\eea{\end{eqnarray}}
\newcommand\de{\mathrm{DE}}

\begin{document}
\title{Why we need to see the dark matter to understand the dark energy}

\author{Martin Kunz}

\address{Physique Th\'eorique, Universit\'e de Gen\`eve, Quai Ernest-Ansermet 24, 1211 Gen\`eve, Switzerland}

\ead{Martin.Kunz@physics.unige.ch}

\begin{abstract}
The cosmological concordance model contains two separate constituents
which interact only gravitationally with themselves and everything else, 
the dark matter and the dark
energy. In the standard dark energy models, the dark matter makes up
some 20\% of the total energy budget today, while the dark energy is
responsible for about 75\%. Here we show that these numbers are only
robust for specific dark energy models and that in general we cannot
measure the abundance of the dark constituents separately without
making strong assumptions.
\end{abstract}

\section{Introduction}

We cosmologists are very proud that our field has finally reached the
status of ``precision science'' over the last decade. The quality of
the current observations of the cosmic microwave background (CMB), the
galaxy distribution and the luminosity distance to type Ia supernovae
(SN-Ia) is indeed impressive, and has allowed the construction of a
concordance model in which the universe contains the known, baryonic,
matter (5\% of the energy density today), radiation (negligible energy
density today), dark matter (20\%) and dark energy (75\%).

The need for dark matter became apparent long ago in order to explain
the motion of galaxies in clusters \cite{darkmat} and the observed
galaxy rotation curves. In cosmology, it is often modelled as a
pressureless fluid with negligible interactions.  Much more recently,
less than ten years ago, new SN-Ia data \cite{sn1a} convinced the
majority of cosmologists that dark energy was needed as well. Until
today the nature of the dark energy is a deep mystery.  Although many
models have been proposed, there are none that can explain its current
abundance in a natural way. The alternative to the model building is a
more phenomenological approach, where one measures the physical
properties of the dark energy. To this end, one introduces a
completely general fluid and tries to determine its characteristics
from observations.

\section{The dark degeneracy}

Unfortunately there is a fundamental problem in this approach. Let us
illustrate this by considering only a perfectly homogeneous and isotropic
universe with vanishing spatial curvature. In this case the line element is
\be
ds^2 = -dt^2 + a(t)^2 dx^2
\ee
with only one degree of freedom, the scale factor $a(t)$, or
equivalently the Hubble parameter $H(t)=\dot{a}/a$. The
energy-momentum tensor has to be compatible with perfect
homogeneity and isotropy, which means that it has to have the
form of a perfect fluid,
\be
T_\mu^\nu = \mathrm{diag} (-\rho(t),p(t),p(t),p(t)) .
\ee
There are two degrees of freedom in the energy momentum
tensor, $\rho(t)$ and $p(t)$. The 0-0 Einstein equation
for this simple example leads to
\be
H^2 = \frac{8\pi G}{3} \rho \label{eq:fried}
\ee
called the Friedmann equation. This equation links the behaviour
of $a(t)$ with the behaviour of $\rho(t)$. The covariant conservation
of the energy momentum tensor gives
\be
\dot{\rho} = -3 H (\rho+p) . \label{eq:cons}
\ee
We find therefore that the pressure $p$ describes the physical
properties of the fluid. Once it is given, then the two equations
can be integrated and solutions for $H$ and $\rho$ are found.
Often cosmologists parametrise the pressure via an auxiliary
function $w$ through an equation of state
\be
p = w \rho .
\ee
Although the equation of state is often written in implicit form
as $p=p(\rho)$ this is not always possible, for example if we are
dealing with one effective fluid composed of two fluids, see
e.g. \cite{phantom} for an explicit example. For this reason
$w$ is often taken to be a free function of time. 

Our original goal of measuring the dark energy properties is now
reduced to the problem of determining the equation of state parameter
of the dark energy, $w_\de$ (the dark matter being characterised
by $w_m=0$). Introducing additionally the quantity $\Omega_m$ for
the relative energy density in matter (both baryonic and dark) today,
and using
the redshift $z$ instead of $t$ as the time variable, it is
easy to combine the equations (\ref{eq:fried}) and (\ref{eq:cons})
and to derive an explicit expression for the equation of state
parameter,
\be
w_\de(z) = \frac{H(z)^2-\frac{2}{3} H(z) H'(z) 
(1+z)}{H_0^2 \Omega_m (1+z)^3-H(z)^2} ,
\label{eq:wh}
\ee 
with $H'=dH/dz$ and $H_0=H(z=0)$. It appears that perfect
knowledge of $H$ implies perfect knowledge of $w$. Unfortunately this
is only true if we also know $\Omega_m$. But cosmological observations
like the luminosity distance only measure $H(z)$, and cannot give any
independent constraints on the matter abundance. We can therefore only
determine a one-parameter family for $w_\de$ as long as we do not have
an independent measurement of $\Omega_m$, e.g. from astroparticle
observations and/or collider measurements.  This has been noticed
before (see e.g.~\cite{old_degpap}) but 
seems to have been forgotten within the community.

\section{Perturbation theory and the CMB}

Our universe is not perfectly isotropic and homogeneous, so is it
possible that this dark degeneracy can be broken by studying e.g.
CMB temperature anisotropy measurements? The short answer is
``no''. The full Einstein equations for our model can be written
as
\be
G_{\mu\nu} = 8\pi G \left( T_{\mu\nu}^{(m)} + T_{\mu\nu}^{(\de)} \right) .
\ee
Clearly the split of the energy-momentum tensor on the right-hand
side is arbitrary. Only the sum of the two tensors is determined
by any measurement involving only gravity (which depends on
the geometry, described by the left-hand side). By allowing a fully
arbitrary $T_{\mu\nu}^{(\de)}$, {\em any} further unknowns cannot
be determined. This concerns not only $\Omega_m$, but also any
possible couplings between the dark quantities, and of course
also possible splits into more than two dark fluids. All these
things {\em cannot} be measured by cosmological probes. If we
(have to) allow for a fully general dark (energy) component,
then for all practical purposes we have to limit ourselves to
a single general dark component. Only non-gravitational measurements
can give us information that goes beyond this description.

It is therefore very worrying that although many papers have been
published putting constraints on $w(z)$, seemingly no-one has found
the degeneracy in their data analysis. There is really no excuse for
groups who have used only distance data. The most likely explanation
is that the parametrisations used were not flexible enough to exhibit
the presence of the full family of solutions. This illustrates once
more that one has to be extremely careful not to put in by hand
what one wants to measure.

However, things become more subtle when also including CMB data, as
now we have to look at perturbation theory.  The theory of general
fluids in first order perturbation theory is well known, see for
example \cite{pt} or \cite{hu}. One finds (in Newtonian gauge,
say) that in addition to $w$ two new variables appear which
characterise the physical properties of the fluid, the pressure
perturbation $\delta p$ and the anisotropic stress $\Pi$. For reasons
of brevity and simplicity we will not discuss the anisotropic stress
here further (and set it to zero), but we cannot avoid taking a closer
look at $\delta p$.

\begin{figure}[ht]
\includegraphics[height=15pc]{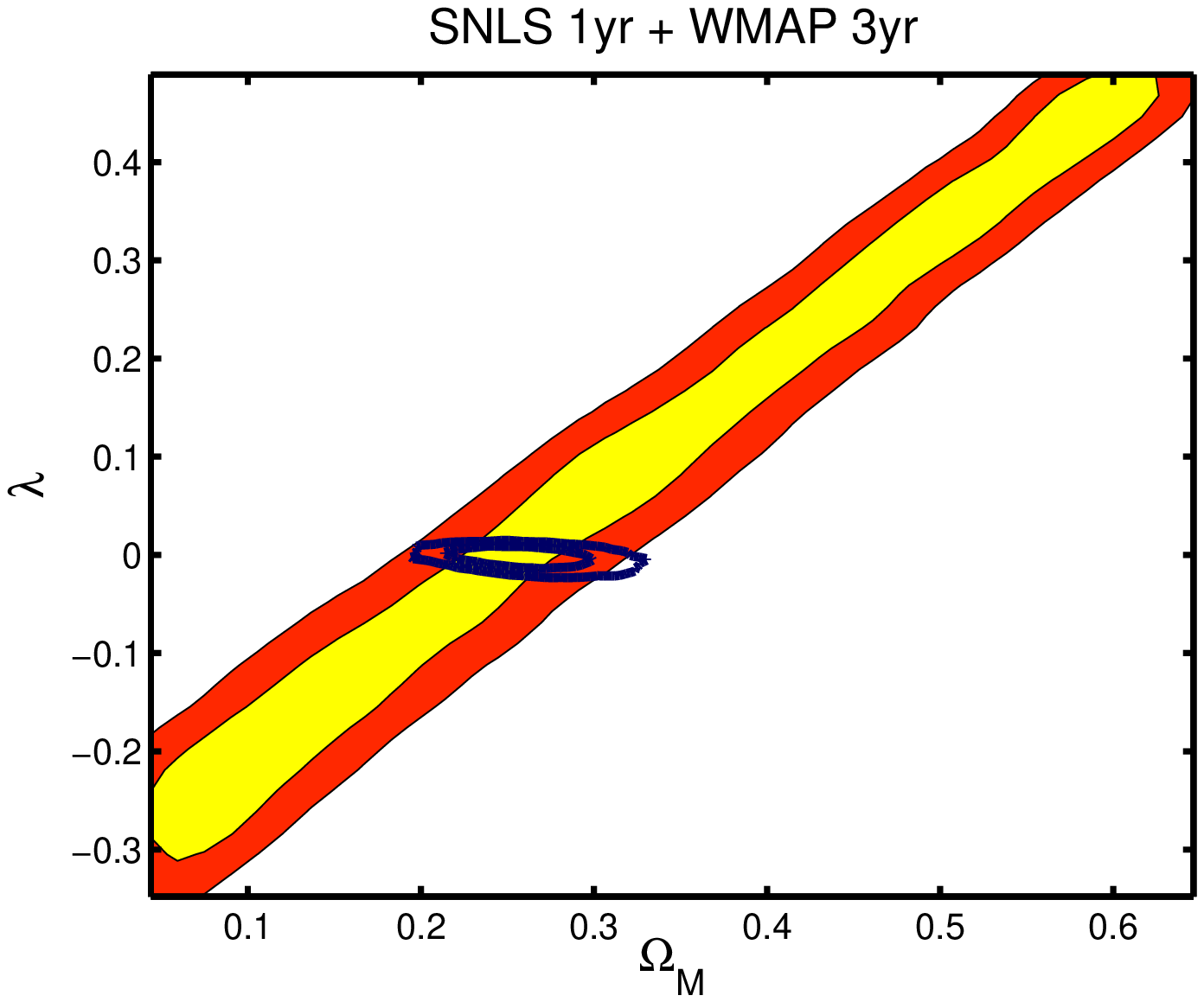}
\includegraphics[height=15pc]{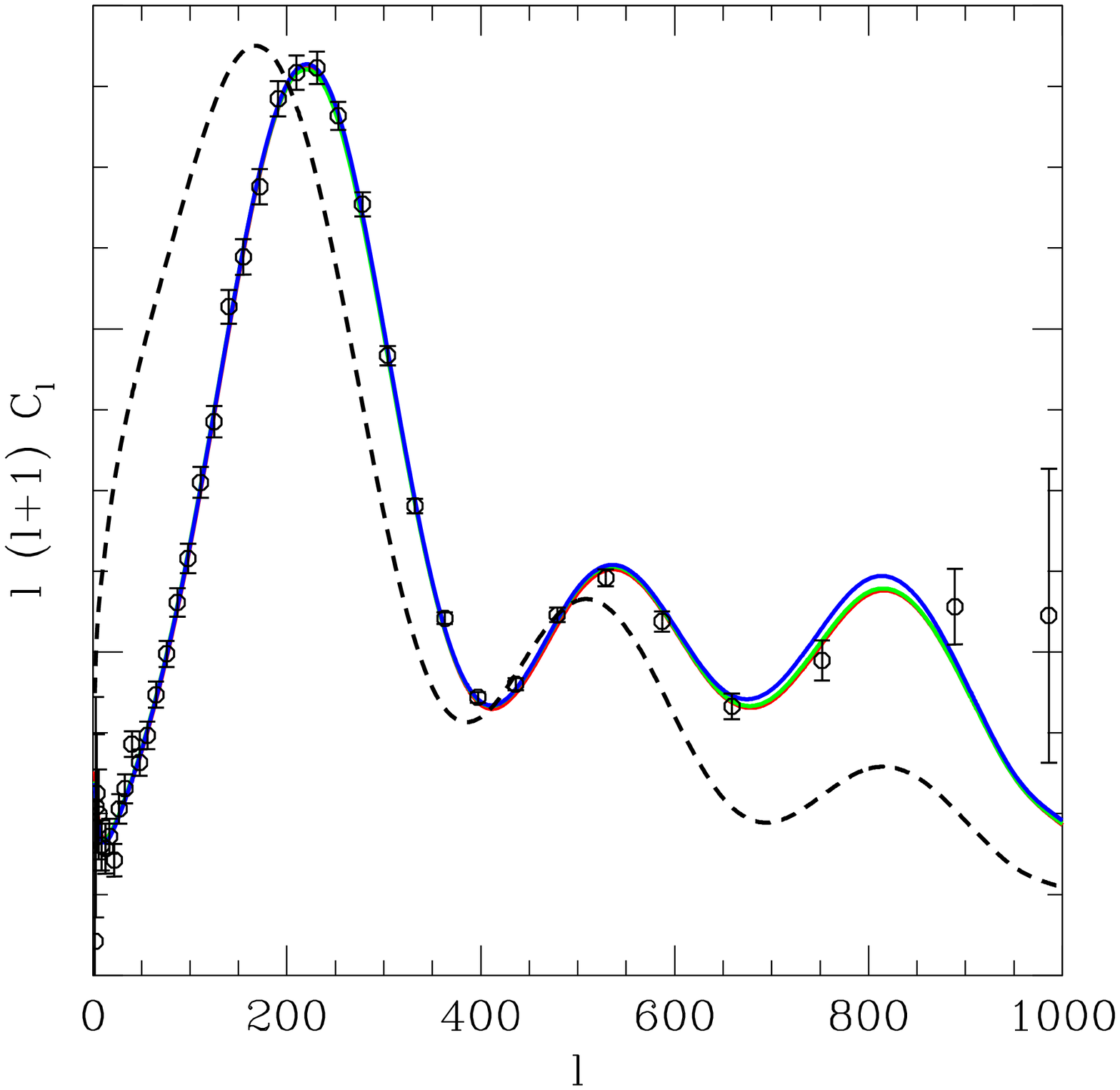}
\caption{\label{fig:cont}CMB and SN-Ia data cannot measure $\Omega_m$
for general dark energy models: In the left panel we plot filled contours 
at $1\sigma$ and $2\sigma$ limits for a model with $\delta p=0$, while the open contours
show the limits for scalar field dark energy. The right panel shows the
corresponding CMB power spectra together with the binned WMAP3 data
(solid curves for $\delta p=0$ and dashed curve for a scalar field model
with $\Omega_m=0.1$).}
\end{figure}

Since we know that a cosmological constant is a good fit to the data
(see e.g. \cite{mydepap}) we construct a family of models that
is degenerate with $\Lambda$CDM. For the equation of state parameter
we use (see \cite{degpap})
\be
w_\de(\lambda ; z) = \frac{-1}{1-\lambda(1+z)^3} .
\ee
The cosmological constant has no perturbations, and
those of the dark matter are characterised by $\delta p = 0$.
Therefore the sum of the two still obey the same condition
we set $\delta p = 0$ overall. Using the WMAP3 CMB data \cite{wmap3} and
the SNLS SN-Ia data \cite{snls} we find the result shown as the
filled contours in the left panel of Fig.~\ref{fig:cont}. Clearly, even using CMB
data, it is impossible to pin down $\Omega_m$. This is shown
explicitely in the right panel of Fig.~\ref{fig:cont} where we show three CMB power spectra
along the degeneracy (solid curves), computed for $\Omega_m = 0.1$,
$0.25$ and $0.6$. They all agree with the binned data and are
practically indistinguishable.

But the most common dark energy model, quintessence, uses a scalar
field. The pressure perturbation of a scalar field is characterised
by $\delta p = \delta \rho$ in its rest frame. This means that the
scalar field dark energy cannot cluster on small scales because of
pressure support. The difference in $\delta p$ makes it possible to
distinguish the two dark components, and now the data allow to
measure $\Omega_m$, as shown by the open contours in Fig.~\ref{fig:cont}
and the dashed CMB power spectrum for $\Omega_m=0.1$ which does
not agree with the data. However, we do not actually {\em know} the
pressure perturbation of the dark energy, nor have we measured it.
We have just fixed it by hand.

\section{Conclusions}

We have seen that cosmology cannot measure separately the properties
of the dark matter and of a general dark energy component. In order to
do that, we either need to impose additional assumptions, for example
that the dark energy is a scalar field, or else we need a
non-gravitational measurement of the dark matter properties,
specifically of its contribution to the total energy density of the
universe. One possibility is a detection of supersymmetry at LHC,
which may in turn determine the abundance and mass of the lightest
stable SUSY particle, one of the best candidates for the dark matter.

As a corollary, if the abundance determined in this way is
not the one expected within the $\Lambda$CDM cosmological concordance
model, one possible explanation is an evolving dark energy.

We can also consider the degeneracy as a test of the
generality of the different approaches to measure the dark energy
equation of state. Since no analysis so far seems to have found it, we
can only wonder what else has been overlooked.

Finally, although we show here that one never can prove experimentally
from cosmological data alone that the dark energy is a cosmological
constant, it is remarkable that a model containing just cold dark
matter and $\Lambda$ fits the data so well. From a model selection
point of view $\Lambda$CDM is still the preferred model because of
its simplicity.

\ack 
It is a pleasure to thank Luca Amendola, Ruth Durrer, Domenico
Sapone and Anze Slosar for interesting discussions. MK acknowledges
funding by the Swiss NSF.

\end{document}